\RequirePackage{fix-cm}
\documentclass[smallextended]{svjour3}       
\smartqed  
\usepackage{graphicx}
\usepackage{float}
\usepackage{amssymb}
\usepackage{amsmath}
\usepackage{subcaption}
\usepackage{endnotes}

\let\footnote=\endnote
\begin{document}

\title{Scaling laws in geo-located Twitter data}
%



\author{Rudy Arthur*         \and
        Hywel T.P. Williams 
}


\institute{Rudy Arthur* corresponding author \at
              Computer Science, University of Exeter \\
              \email{R.arthur@exeter.ac.uk}           
           \and
           Hywel Williams \at
              Computer Science, University of Exeter    \\ 
\email{H.T.P.Williams@exeter.ac.uk}
}

\date{Received: date / Accepted: date}

\maketitle

\begin{abstract}
We observe and report on a systematic relationship between population density and Twitter use. Number of tweets, number of users and population per unit area are related by power laws, with exponents greater than one, that are consistent with each other and across a range of spatial scales. This implies that population density can accurately predict Twitter activity. Furthermore this trend can be used to identify `anomalous' areas that deviate from the trend. Analysis of geo-tagged and place-tagged tweets show that geo-tagged tweets are different with respect to user type and content. Our findings have implications for the spatial analysis of Twitter data and for understanding demographic biases in the Twitter user base.
\end{abstract}

\section{Introduction} \label{sec:introduction}

Twitter is a social media platform whose open access API makes it very popular with researchers. Twitter data has been used to study earthquakes \cite{Sakaki:2010}, wildfires \cite{Boulton:2016}, floods \cite{Arthur:2017}, language \cite{Eisenstein:2010}, land use patterns  \cite{Frias:2012}, public health \cite{Ghosh:2013}, happiness \cite{Mitchell:2013} and many other topics. Twitter, with around 328 million users as of August 2017, \footnote{https://www.statista.com/statistics/272014/global-social-networks-ranked-by-number-of-users/} is now established as a key data source in quantitative social science and for geographic applications such as event detection or regional comparisons, with some suggesting that it is perhaps even overused \cite{Tufekci:2014}. Given this, it is important to understand demographic biases like age, gender, ethnicity and to understand the relationship between population density and volume of Twitter activity, in order to normalise event detection algorithms. A lot of work has been done on this topic already \cite{Mislove:2011} \cite{Hecht:2014} \cite{Longley:2015} \cite{Malik:2015}. 

Studies of demographic bias find users in cities and urban areas over-represented in collections of Twitter data. Mislove et. al. \cite{Mislove:2011} found ``...Twitter users are more likely to live within populous counties than would be expected from the Census data, and that sparsely populated regions of the US are significantly underrepresented.'' Hecht et. al. \cite{Hecht:2014} find that urban users are over-represented in geo-located social media, and also provide more information than rural users. Longley et. al. \cite{Longley:2015} find Twitter in London is not representative of the true age profile or gender ratio but representation of ethnicity is more reflective of the true population. They also find a fairly strong level of ethnic segregation among Twitter users in London. Recent work by Malik et. al. \cite{Malik:2015} on US Twitter usage has identified other demographic variables such as higher median income, being in an urban area or having more young people as being predictive of more geolocated tweets originating from an area. However, to our knowledge, no one has studied the relationship between population density and Twitter activity.

It is the aim of this paper to determine whether population density affects Twitter usage in a systematic way. In particular we are motivated by work of Takhteyev et. al. \cite{Takhteyev:2012} and Stephens et. al. \cite{Stephens:2014} which shows that a user's connections are not randomly distributed around the globe, but rather Twitter builds on existing social structures (e.g. neighbourhoods, cities, airline connections, languages) and does not supersede them. The work of Tizzoni et. al. \cite{Tizzoni:2014} is very interesting in this regard. In building a model of human interactions as a reaction-diffusion process on a graph, they construct a graph of Twitter users, using mentions to connect users in the same metropolitan area. They find that the connectivity (rescaled cumulative degree, the total number of links times the proportion of the population who are Twitter users) scales super-linearly with population size. 

The search for systematic effects with population density and urban bias of Twitter usage relates closely to work on scaling laws in cities  \cite{Bettencourt:2007} \cite{Bettencourt:2010} \cite{Bettencourt:2013} \cite{Batty:2008}. Scaling laws are power law relationships that describe the variation in some quantity with population, population density or some other size metric. For example the amount of food consumed is a linear function of the population, while the amount of electrical cable is a sub-linear function of population density (since denser neighbourhoods require fewer connections to service the same number of people). Many creative or social outputs like wages or number of patents scale super-linearly with city size. The work of \cite{Schlapfer:2014}, similarly to \cite{Tizzoni:2014}, showed that social networks (this time constructed from mobile phone data) also scale super-linearly with city size.

The outline of this paper is as follows: First we calculate three key statistics of an area: 
\begin{itemize}
\item $T$, tweet density, number of tweets/km${}^2$
\item $U$, user density, number of users/km${}^2$ 
\item $P$, population density, population/km${}^2$. 
\end{itemize}
We then show that these quantities are related by power laws, with exponents that are constant across a range of spatial scales. The very consistent scaling we observe allows prediction of Twitter activity as a function of population density. We can then account for systematic urban biases and rank locations based on their deviation from scaling, which we propose is a better way to identify places with anomalous Twitter behaviour than simple statistics like number of tweets or users per capita.

\section{Methods}
\subsection{Tweet collection}

We collected tweets from the South-West UK by passing a geographical bounding box with longitude from -5.8 to -1.2 and latitude from 49.9 to 52.2 to the Twitter Streaming API. This area is reasonably representative of the UK, having a population of around 9.6 million people and containing a mixture of rural and industrial areas, as well as large cities (Bristol, Southampton, Cardiff). The collection ran from 11/4/2016 to 1/10/2017 with some gaps in the collection (due to machine downtime) which cover all of December 2016 and all of March 2017. The Twitter Streaming API is free of cost and easy to access compared to the full Twitter datastream. This makes it the usual choice for research purposes. The Streaming API is rate limited, only allowing us to collect a restricted number of tweets, at most 1\% of all the tweets on Twitter \cite{Morstatter:2013}. Since we are looking at a rather small area, and only at tweets with geo-location tags, the threshold for rate limiting will only rarely, if ever, be crossed and the Streaming API should be sufficient for a general survey of tweets. 

We use two metadata properties of each tweet to locate it: geo-tags and place-tags. Geo-tags are GPS co-ordinates added by the user's mobile device that give a precise location for a particular tweet. According to the Twitter API documentation \footnote{https://dev.twitter.com/overview/api/places  (Accessed October 2017)}: ``Places are specific, named locations with corresponding geo coordinates. They can be attached to tweets by specifying a place\_id when tweeting. Tweets associated with places are not necessarily issued from that location but could also potentially be about that location.'' Place-tags are often quite precise and all of our tweets have a populated ``place'' field. Place-tags are added to the tweets of a user who opts in to using Twitter's location services. Once a user opts in to location services, e.g. by tagging a tweet in a certain place, all subsequent tweets will automatically include a general location label \footnote{https://support.twitter.com/articles/78525\#  (Accessed October 2017)} as a place-tag. 

We will show geo-tagged tweets are a minority (as other studies have found e.g. \cite{Leetaru:2013}) and have different statistics than place-tagged tweets. Other studies have looked at how geo-tagged tweets differ from non-geotagged tweets,   e.g. Pavalanathan et. al. \cite{Pavalanathan:2015} found GPS-tagged tweets are written more often by young people and women, use more geographically specific words and are generally longer. There is also an extensive literature on the problem of geo-locating users based on inference from the user's location field, words in their tweets or by locating based on their friends' locations \cite{jurgens:2013} \cite{schulz:2013} \cite{ajao:2015}, \cite{compton:2014}. However all of our tweets have a `place' tag, provided by the API, based on Twitter's own location services. Since Twitter is able to access more information about each tweet than we can get from the API (e.g. GPS, cell tower signal or data about nearby wireless access points \footnote{https://support.twitter.com/articles/118492\#  (Accessed October 2017)}) the location information provided in the place field is likely to be of good quality and we do not use any location inference methods.

\subsection{Creating a grid}

We divide the geographical bounding box into a $X \times X$ grid. Since our original bounding box is not square, these grid cells are rectangular. As some grid boxes are over coastal areas, not all grid boxes cover the same land area. Let $A$ be the land area of a grid box. To calculate $A$ the shape-files for the UK were obtained from the GADM database of Global Administrative Areas \footnote{gadm.org}. $A$ is calculated by projecting the UK polygon from WGS84 co-ordinates (based on the curvature of the earth) to an equal area projection and using standard planar techniques for calculating polygonal area.

Our choice of grid is arbitrary, we could just as easily have covered the collection area with hexagons or some other shape. We will show that our results are robust across a range of grid sizes, so that we actually have a scaling `law' holding over a certain distance scale that does not depend on how the areas are chosen.

\subsection{Measuring tweet and user density}

Let $N_t$ be the number of tweets in a grid box, $N_u$ the number of users in a grid box, $N_p$ the population. We define
\begin{align}
T &= N_t / A \\ \nonumber
U &= N_u / A \\ \nonumber
P &= N_p / A \nonumber
\end{align}
where $T$ is the number of tweets/km${}^2$, $U$ is the number of users/km${}^2$ and $P$ is population/km${}^2$.

Each tweet is located, via the geo-tag or place tag, to either a bounding box given in the `place' metadata or a single point given by a zero area bounding box or a GPS co-ordinate. When the tweet is located at a single point we add 1 to the count for the grid box containing that point. When the tweet is localised in a bounding box we add to every overlapping grid box, indexed by $b$, a fractional value:
\begin{equation}
f_{jb} = \frac{\text{Area(bounding box}_j \cap \text{grid box}_b \text{)} }{\text{Area(bounding box}_j \text{)}}
\end{equation}
to the count of tweets in that box, where $j$ labels the tweet. 

To count the users in a grid box we first count the number of tweets per user. For a user $i$ let $N_t(i)$ be the number of tweets by that user. Since users can post from multiple grid boxes we split the contribution of each user proportionally across the places they tweet from. So, for each of user $i$'s tweets, labelled $j$, we calculate $f_{jb}$ as above, for all overlapping grid boxes $b$. We add $\frac{f_{jb}}{N_t(i)}$ to the count of number of users in $b$. Thus if a user only ever tweets from within a single grid box we will end up adding $1$ to the count of users in that grid box, if a user divides their time between two locations equally we will add $0.5$ to the count of users in each grid box, and so on. 

To measure the population in a grid box we used the latest UK mid-year population estimate\footnote{https://www.ons.gov.uk/peoplepopulationandcommunity/populationandmigration/populationestimates (Accessed October 2017)} in each Lower Super Output Area (LSOA). LSOAs are polygonal areas designed by the Office for National Statistics to improve the reporting of small area statistics in the UK. They contain at least 1000 inhabitants with a mean of 1500 and are designed to be as consistent as possible in population size. They range in area from very small (smallest is $\sim 0.05$ km${}^2$) to very large (largest is $\sim 250$ km${}^2$) as we move from cities to the countryside. We downloaded the LSOA polygons from the UK Data Service\footnote{https://borders.ukdataservice.ac.uk/easy\_download.html (Accessed October 2017)}. When an LSOA intersects multiple grid boxes we divide the population in the LSOA among the intersecting boxes proportionally to the area of the intersection with the LSOA. When multiple LSOAs lie in the same grid-box we sum the intersecting LSOA populations. Around $80\%$ of LSOAs are smaller than our smallest grid box ($\sim 5$ km${}^2$). Although the population within a LSOA is not necessarily uniformly distributed, the LSOAs are typically small enough that this is not a concern for our analysis, though it may result in some minor artefacts in rural areas when using very small grid boxes.

\section{Results}

We collected 26631472 tweets in total during our collection period. Of these there were 2500296 geo-tagged with GPS co-ordinates inside the target area and 14210527 place-tagged with bounding boxes contained in the target area. The remaining tweets either had geo-tags outside the target area or place-tag bounding boxes not fully contained inside. There were 140026 users who made at least 1 geo-tagged tweet and 311843 users with at least one place-tagged tweet, with some overlap between these sets. When processing the data, if a tweet is geo-tagged we use that as the tweet location. If not we check the place\_type field. If it is `country' (e.g. United Kingdom) or `admin' (e.g. South-West), we discard it since it does not provide sufficient precision. Otherwise we use the bounding box associated with that tweet for the location.

We also removed 6 very active automated accounts (bots) whose tweets make up more than 1\% of the total number of tweets in our dataset. These accounts are linked to e.g. automatic weather stations, solar panels, or phone number services that produce a huge number of geo-tagged tweets. These 6 users produced a combined 901121 geo-tagged tweets. We remove these users as extreme outliers in our data set leaving 1599175 geo-tagged tweets. The remaining data comes from a mix of individual users and organization accounts. 

We make the ansatze:
\begin{align}\label{eqn:ans}
T = CU^\gamma \\ \nonumber
U = BP^\beta \\ \nonumber
T = AP^\alpha \nonumber
\end{align}
A naive model might assume all people are equally likely to become Twitter users and all users are equally likely to author a tweet in a given time period. This would make the number of users directly proportional to the population (so $\beta = 1$) and number of tweets per user a constant (so $\gamma = 1$). Deviations from these values tell us about the actual adoption rates of Twitter and the behaviour of Twitter users. We find the actual values for $\gamma, \beta, \alpha$ by fitting to our data. 

If population density is the only relevant variable then by consistency we should have:
\begin{equation}\label{eqn:consistency}
\alpha = \beta \gamma.
\end{equation}
Deviations from this relation imply other variables are important for prediction of Twitter adoption and Twitter use in an area.

\subsection{Place-tagged tweets}

\begin{figure}[H]
\centering 
\includegraphics[width=\textwidth]{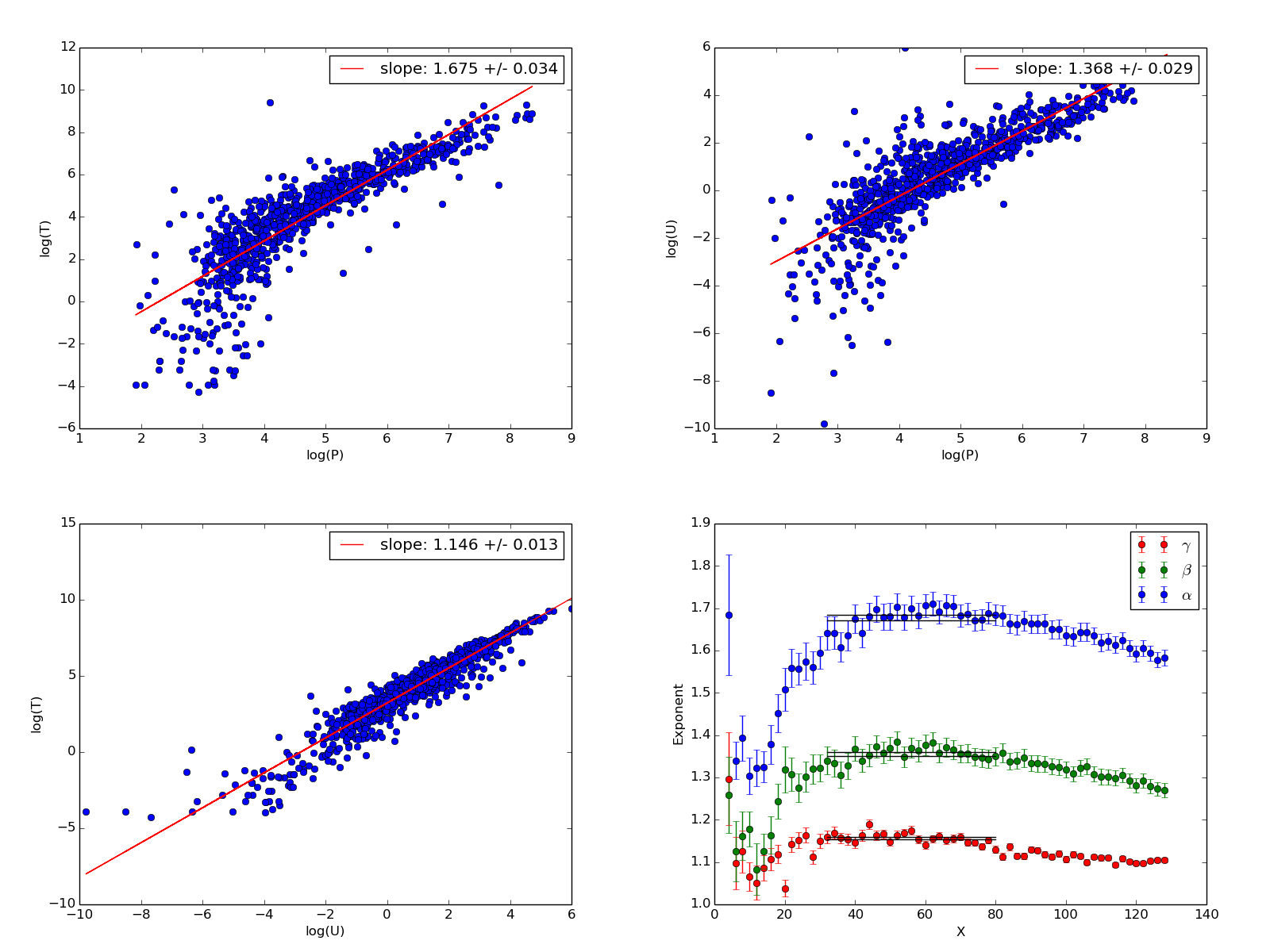}

\caption{Scaling laws for place-tagged tweets. Top left: Tweet density $T$ versus population density $P$, slope of the line is $\alpha$. Top right: User density $U$ versus population density $P$, slope of the line is $\beta$. Bottom left: Tweet density $T$ versus user density $U$, slope of the line is $\gamma$. Values are taken from a $40 \times 40$ grid. Bottom right: Exponents $\alpha$, $\beta$, $\gamma$ versus grid size $X$. The horizontal lines mark the largest range of grid sizes across which the exponents agree within errors, showing the average value of the exponent across the range.}
\label{fig:place_plots}

\end{figure}

Figure \ref{fig:place_plots} shows fits to Equation \ref{eqn:ans} for a $40 \times 40$ grid, using place tags to locate tweets and users. Each point corresponds to a grid box with at least one tweet and one resident. All of the exponents are significantly greater than one. The bottom right panel shows a range of different grid sizes (finer grids on the right) demonstrating that the exponents are consistent for a range of grid sizes, approximately $X=32$ to $X=80$, corresponding to physical sizes roughly 82km${}^2$ to 13km${}^2$. We will call this the `scaling window'. The plot of tweet density, $T$, versus population density, $P$, in Figure \ref{fig:place_plots} shows the biggest deviations from power-law behaviour at low population density, perhaps indicating that at low population density other factors determine the number of tweets and users.

\subsection{Geo-tagged tweets}
\begin{figure}[H]
\centering 

\includegraphics[width=\textwidth]{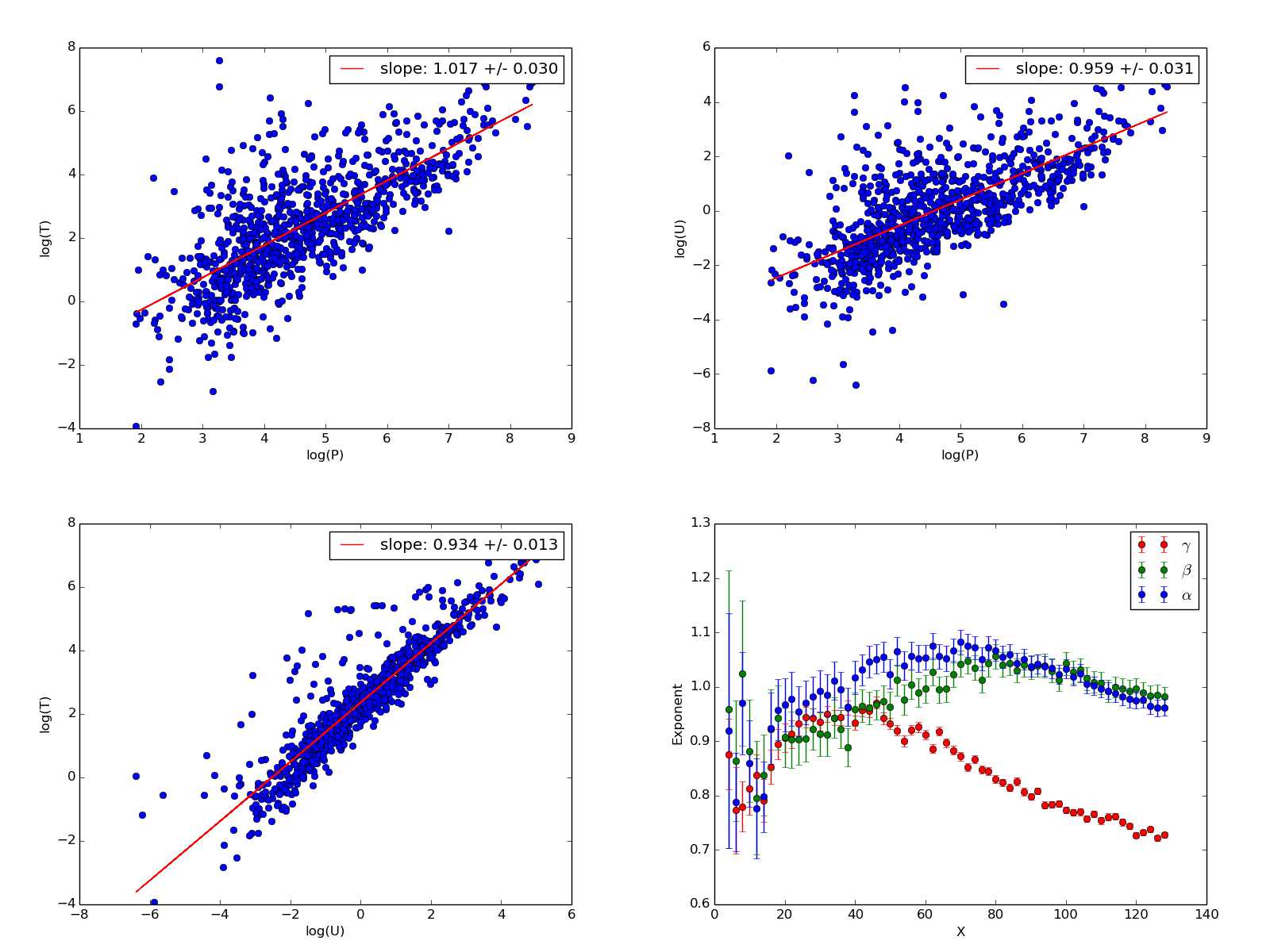}

\caption{Scaling laws for geo-tagged tweets. Top left: Tweet density $T$ versus population density $P$, slope of the line is $\alpha$. Top right: User density $U$ versus population density $P$, slope of the line is $\beta$. Bottom left: Tweet density $T$ versus user density $U$, slope of the line is $\gamma$. Values are taken from a $40 \times 40$ grid. Bottom right: Exponents $\alpha$, $\beta$, $\gamma$ versus grid size $X$. }
\label{fig:gps_plots}

\end{figure}
Figure \ref{fig:gps_plots} shows fits to Equation \ref{eqn:ans} for a $40 \times 40$ grid, using geo-tags to locate tweets and users. Each point corresponds to a grid box with at least one GPS tagged tweet and one resident. The exponents $\alpha$ and $\beta$ are consistent with one, while the exponent $\gamma$ is less than one. The bottom right panel shows a range of different grid sizes. The exponents, especially $\gamma$, vary a lot as the spatial scale changes. There is a small range of grid sizes, X=40 to X=64, where the exponents are approximately constant (corresponding to physical sizes roughly 52km${}^2$ to 20km${}^2$). However this relationship does not seem to be as robust as that found for place-tagged tweets. Most importantly, though we find reasonably accurate fits to the data, the exponents are all significantly smaller, with the tweets per user exponent, $\gamma$, less than 1. These facts indicate that geo-tagged tweets are markedly different than place-tagged ones. We will discuss possible reasons for this in the following section.

\section{Analysis}

\subsection{Source of tweets}

\begin{table}
\centering
\begin{tabular}{|c|c|c|}
\hline
Rank & Source & Proportion \% \\ \hline
1 & Instagram & 60.6 \\ \hline
2 & Sandaysoft Cumulus & 5.4 \\ \hline
3 & dlvr.it & 4.4 \\ \hline
4 & Foursquare & 4.3 \\ \hline
5 & dlvrit.com & 3.4 \\ \hline
\end{tabular}
\caption{Top 5 sources of geo-tagged tweets in our data set. 1599175 tweets in total. } \label{tab:geo_source}
\end{table}

\begin{table}
\centering
\begin{tabular}{|c|c|c|}
\hline
Rank & Source & Proportion \% \\ \hline
1 & Twitter for iPhone & 60.6 \\ \hline
2 & Twitter for Android & 21.9 \\ \hline
3 & Twitter Web Client & 12.5 \\ \hline
4 & Twitter for iPad & 3.5 \\ \hline
5 & Tweetbot for iOS & 1.1 \\ \hline
\end{tabular}
\caption{Top 5 sources of place-tagged tweets in our data set. 14210527 tweets in total. } \label{tab:place_source}
\end{table}

Twitter provides a field in the meta-data associated with each tweet: `source'. This records the utility or application that was used to post the tweet e.g. `Twitter for Android'. For tweets with geo-tags and tweets with only place-tags we rank the most common sources in Tables \ref{tab:geo_source} and \ref{tab:place_source}. These two tables can explain much of the difference between geo-tagged and place-tagged tweets.

Looking at Tables \ref{tab:geo_source} and \ref{tab:place_source}, we see very different sources for the different types of tweet. The majority of tweets with geo-tags are Instagram posts that have been shared on Twitter. Sandaysoft Cumulus\footnote{http://sandaysoft.com/products/cumulus} is software for personal weather stations, dlvrit\footnote{https://dlvrit.com/} is an automated social media service for marketers and Foursquare\footnote{https://foursquare.com/} is a social media platform for consumer recommendations. Thus geo-tagged posts are predominantly shared posts from other websites (Instagram and Foursquare), automated weather bots and marketing accounts. In contrast, place-tagged posts typically originate from Twitter clients on mobile devices and the Twitter website itself. 

Instagram is itself a very active and popular social media platform, making it likely that users would respond to Instagram posts directly on Instagram, rather than via Twitter. It is also unlikely that bots (of the kinds found here) will reply to other bots. To check this we can examine the tweet meta-data fields `in\_reply\_to\_status\_id', `in\_reply\_to\_user\_id' and `quoted\_status\_id'. If either of the first two fields are non-empty the tweet is a reply to another user, if `quoted\_status\_id' is non-empty the tweet quotes another tweet. Twitter also has `retweets'; however, requesting tweets by location from the API, as we did, returns no retweets, presumably as the retweeter's location may be different than the location of the original tweet. Thus we can only see replies and quotes in our dataset.
Of the 1599659 geo-tagged Tweets, 85575 are replies and 3051 are quotes, making up 5.5\% of the total. In contrast, for the 14215471 place tweets there are 5783349 replies and 1455559 quotes, making up 50.9\% of the total. 

Clearly tweets with geo-tags are qualitatively different from tweets with place-tags, in both the kind of user they originate from and the content they contain. Since geo-tagged tweets are mostly either from bots or from users sharing social media content from other platforms, these tweets do not seem to represent the kind of Twitter-native human social interaction that researchers appear to be looking for when they study Twitter. From here on we will only consider place-tagged tweets, since these seem to better represent the kind of Twitter-native data that most researchers intend to study.
 
\subsection{Checking the consistency relation}

\begin{figure}[H]
\centering 

\includegraphics[width=\textwidth]{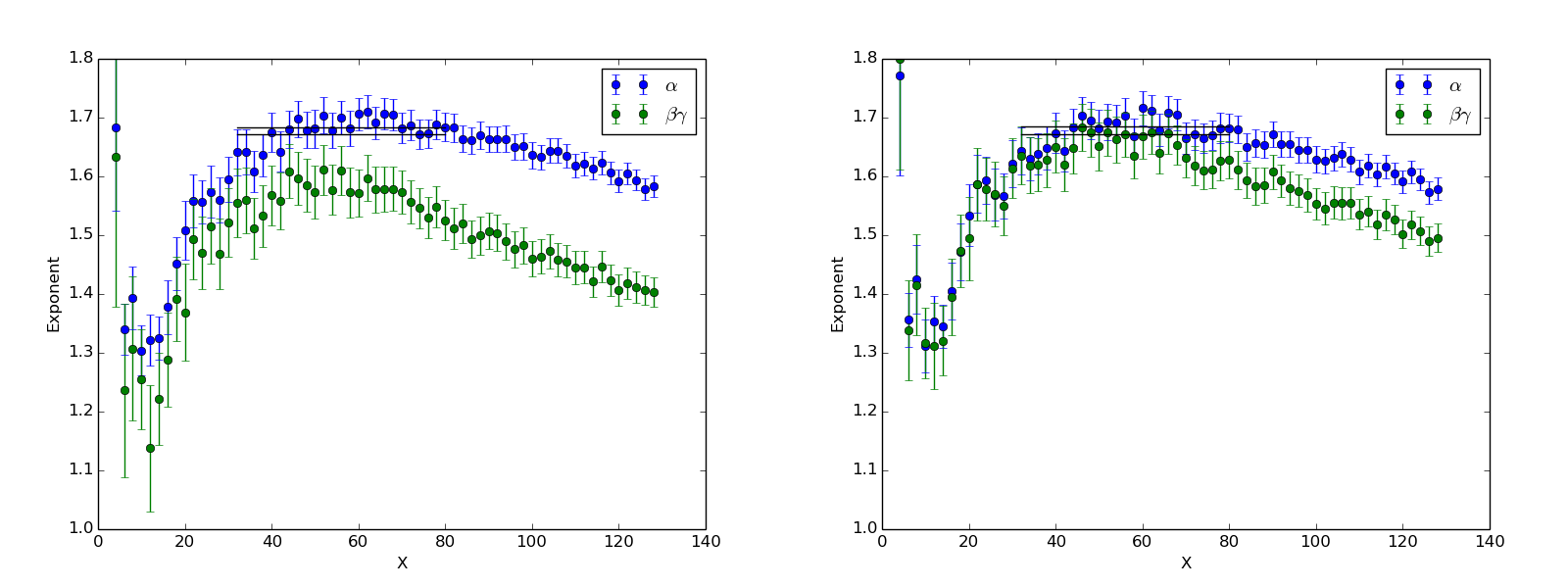}

\caption{Left: Comparing the exponent $\alpha$ obtained from fitting the data to $\beta \gamma$ using all the place-tagged tweets. Right: Comparing the exponent $\alpha$ obtained from fitting the data to $\beta \gamma$ using only users with at least 10 place-tagged tweets during the data collection period. Black line is the average of $\alpha$ across grid sizes from $X=32$ to $X=80$.}
\label{fig:consistency_plots}

\end{figure}

We check the consistency relation give in Equation \ref{eqn:consistency} in Figure \ref{fig:consistency_plots}. We find some disagreement, indicating that population density is not the only factor accounting for the number of tweets per user. We get better agreement by looking only at users who have tweeted multiple times in our observation region over the course of the study. For example, Equation \ref{eqn:consistency} is more closely satisfied if we restrict our fits to users who made at least 10 tweets during the whole observation period. This reduces our count of 14210527 tweets from 311843 users to 13649169 tweets from 114565 users. Since it is more likely that users who tweet often in a region live there, and so are counted in the census, the improved agreement is to be expected. The lack of exact agreement is not surprising; multiple studies \cite{Hecht:2014} \cite{Malik:2015} have shown that gender, race, age and other demographic variables are predictive of Twitter use, which are not captured here. Indeed it is surprising that population density alone does such a good job of predicting user density and tweet density.

\subsection{Deviations from scaling relationships}

\begin{figure}[H]
\centering 
\includegraphics[width=\textwidth]{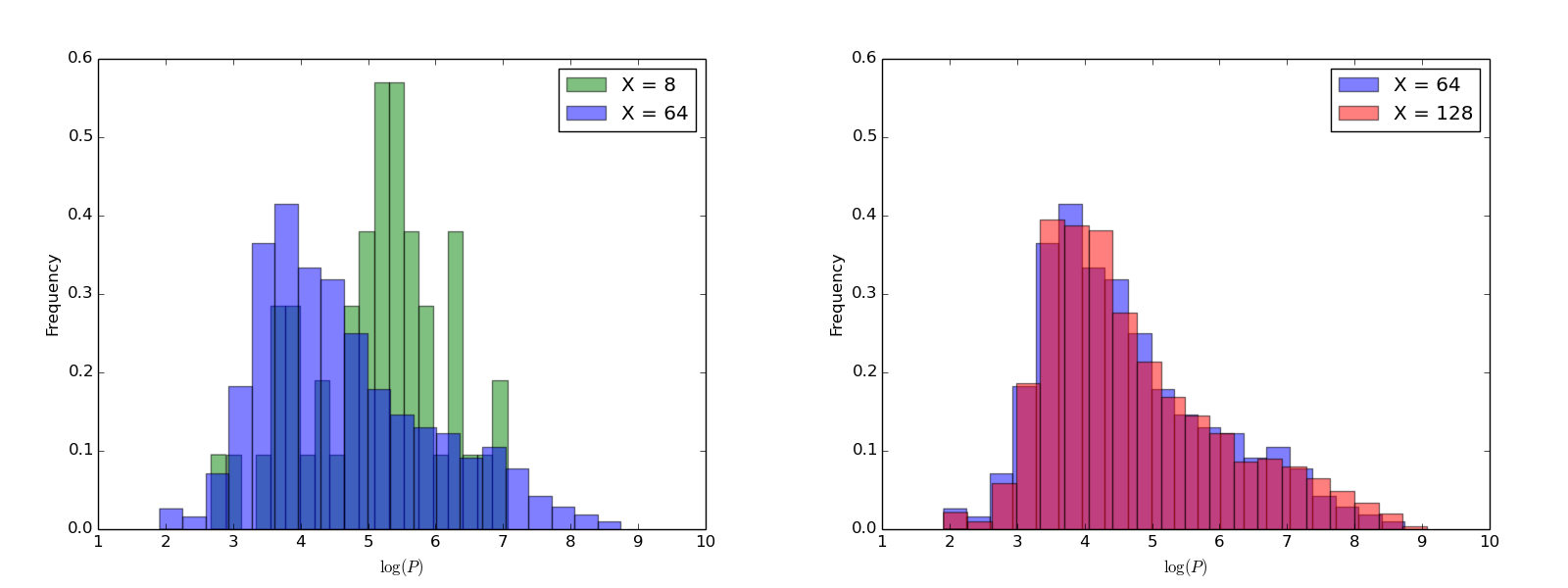}

\caption{Histogram of $\log(P)$. Comparing an $8 \times 8$ grid and a $64 \times 64$ grid on the left and a $64 \times 64$ with a $128 \times 128$ grid on the right. The $128 \times 128$ and $64 \times 64$ grids give similar distributions with long tails while the $8 \times 8$ grid is concentrated around a central value.}
\label{fig:hist_plot}

\end{figure}

For large grid boxes (e.g. $X < 32$), the exponents fall off rapidly, Figure \ref{fig:place_plots}. This is because using very large grid boxes does not generate a diverse sample of homogeneous grid boxes that each represent a different area class (e.g. rural, suburban, urban). Instead large grid boxes are likely to capture a heterogeneous mixture of rural and urban areas and average them together. Smaller boxes are more likely to capture a homogenous area (e.g. rural or urban, but not both). This effect can be seen by plotting a histogram of population density for large, intermediate and small grid boxes, Figure \ref{fig:hist_plot}. We see that the distribution of $P$ over large areas (e.g. $8\times 8$) is qualitatively different to the distribution over smaller areas (e.g. $64\times 64$, $128\times 128$).

\begin{figure}[H]
\centering 

\includegraphics[width=\textwidth]{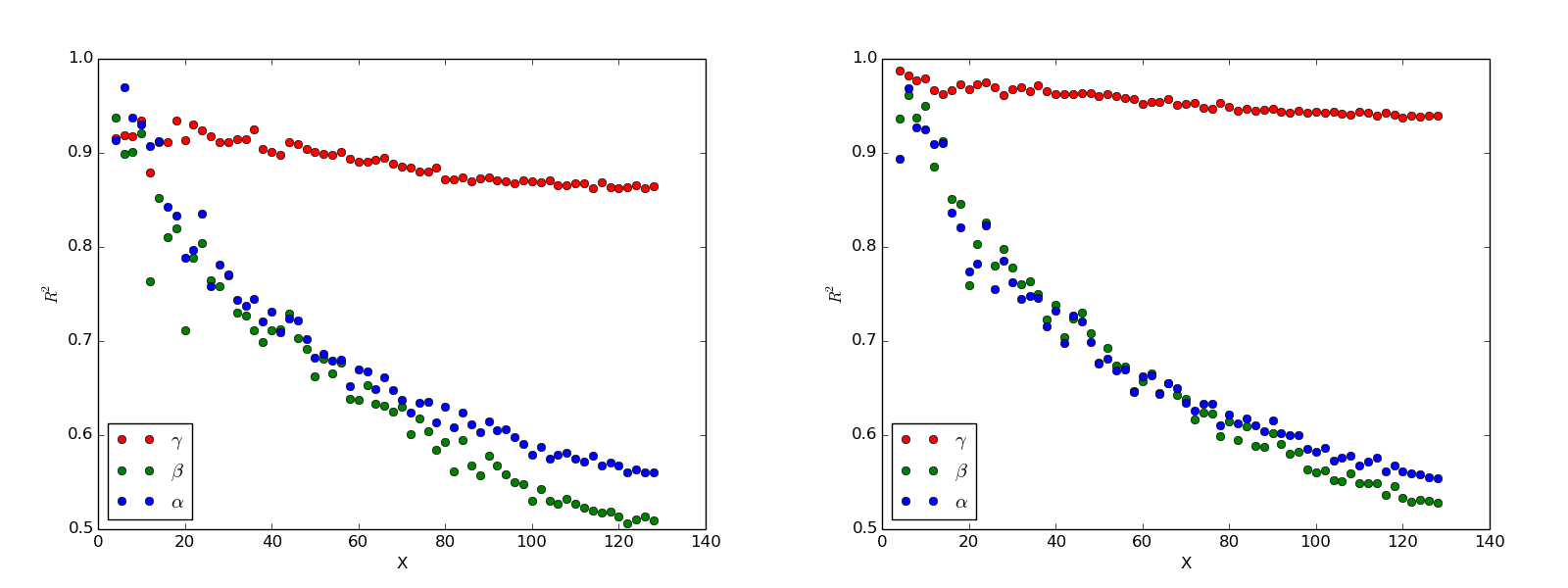}

\caption{Coefficient of determination $R^2$ for each fit of the data to Equation \ref{eqn:ans} across a range of grid sizes. Left: Fits using all place-tagged tweets. Right: Restricting the fits to users with at least 10 tweets. Tweet density $T$ is very well predicted by user density $U$. Across the scaling window from $X=32$ to $X=80$, between 75\% and 60\% of the variance in $T$ and $U$ is predicted by population density $P$.}
\label{fig:fitquality}
\end{figure}

Exponents also fall away for small grid boxes, $X>80$. This is because very small boxes may contain too few people to be treated as `populations'. We see this reflected in a decreasing fit quality as the grid resolution increases. Figure \ref{fig:fitquality} shows the coefficient of determination, $R^2$, for each fit on each grid. $R^2$ is the proportion of the variance in the dependent variable that is predictable from the independent variable. For predicting $T$ given $U$ we get $R^2$ greater than $0.9$ for most grid sizes. The $R^2$ value falls off more rapidly with decreasing grid box sizes for fits of $T$ and $U$ against $P$. As grid boxes get smaller, local and historical explanations are necessary to predict Twitter usage. In the limit, when the boxes become small enough to contain only a handful of users, clearly a model based on the characteristics of individuals rather than areas is more appropriate. This is reflected in the decreasing $R^2$ for the $\alpha$ and $\beta$ fits and the deviation from scaling for finer grids, $X>80$. 

\section{Finding Anomalies}

\subsection{Measuring anomalies}

Given the robust trends observed in the place-tagged data we can predict the typical Twitter activity for an area fairly accurately, given its population density or the density of Twitter users. We can then ask about places which deviate most strongly from the observed trend. It is this deviation from the trend, rather than unusually high or low numbers of tweets, that one should use to identify an anomalous area. An area might have a very high or very low number of tweets, but if this number is well-predicted by Equation \ref{eqn:ans}, then this area is `typical'. If it has many more or less than expected we might inquire further into the possible causes.

We will call the difference between predicted tweet density $T = CU^\gamma$ and measured tweet density $\tilde{T}$ the `anomaly':
\begin{equation}
A_{TU} =  \tilde{T}  -  CU^\gamma 
\end{equation}
Large positive values of $A_{TU}$ indicate many more tweets than expected, large negative values indicate many fewer. We could define the other anomalies $A_{TP}$ and $A_{UP}$ similarly. However, examining Figures \ref{fig:place_plots} and \ref{fig:fitquality} we see that fits of $T$ against $U$ are most precise. Until we have a more precise model for $U$ and $T$ as a function of $P$ the anomalies $A_{TP}$ and $A_{UP}$ would be likely to simply measure fitting error. For $T$ versus $U$ the simple power law fits the data very well across all scales, so $A_{TU}$ is measuring something significant about the area. 

We can also normalise the anomaly:
\begin{equation}
\hat{A}_{TU} = \frac{  \tilde{T}  - CU^\gamma }{ \sqrt{ CU^\gamma \tilde{T}} }
\end{equation}
and measure relative deviation from the fit instead of absolute deviation. The absolute anomaly will tend to emphasise cities and towns, since they generate more tweets and have proportionally larger deviations, whereas the relative anomaly $\hat{A}_{TU}$ controls for population density and puts all areas on an equal footing. Relative deviation will classify a rural area with 4 tweets observed where 2 tweets were expected as equivalently anomalous to a town with 40000 tweets observed where 20000 tweets were expected. 

\subsection{Mapping anomalies}

\begin{figure}
\centering 
 \includegraphics[width=\textwidth]{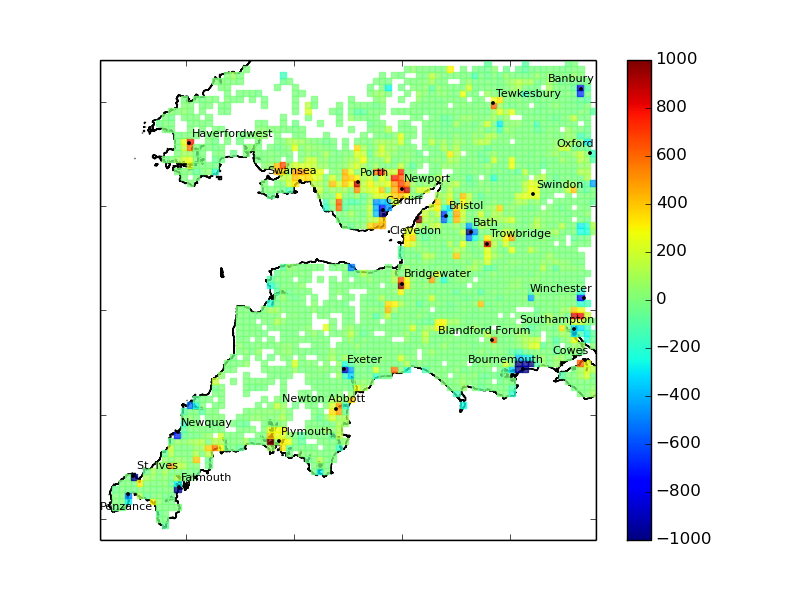}

\caption{ $A_{TU}$  plotted as a heatmap over the South-West UK (with $|A_{TU}|$ capped at 1000 for better resolution). White space indicates less than one tweet or less than one person per km${}^2$ in the grid box.}
\label{fig:anomaly_plots}

\end{figure}
\begin{figure}
\centering 
 \includegraphics[width=\textwidth]{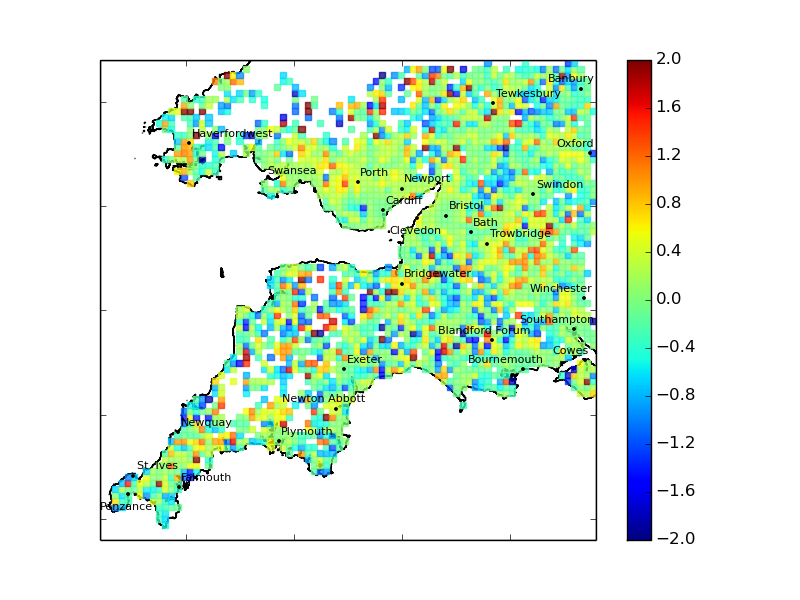}

\caption{ $\hat{A}_{TU}$  plotted as a heatmap over the South-West UK (with $|\hat{A}_{TU}|$ capped at 2 for better resolution). White space indicates less than one tweet or less than one person per km${}^2$ in the grid box.}
\label{fig:rel_anomaly_plots}

\end{figure}

Using the place-tagged tweets from users who tweeted at least 10 times during the collection period, Figure \ref{fig:anomaly_plots} shows $A_{TU}$ and Figure \ref{fig:rel_anomaly_plots} shows $\hat{A}_{TU}$. The absolute anomaly $A_{TU}$ will emphasise towns and cities. In Figure \ref{fig:anomaly_plots} we see some towns with deficits (negative anomalies) like Exeter, Bournemouth and Cardiff and some with excesses (positive anomalies) like Newport, Plymouth and Southampton. Looking at the relative anomaly $\hat{A}_{TU}$ there is no obvious pattern of positive or negative relative anomalies associated with the countryside or towns and cities, i.e. we have successfully detrended the tweet density data. Local hotspots might be examined to discover their causes (e.g. tourist attractions or festivals), but we do not attempt this here.

\subsection{(Not) Explaining anomalies with youth}

\begin{figure}
\centering 
\includegraphics[width=\textwidth]{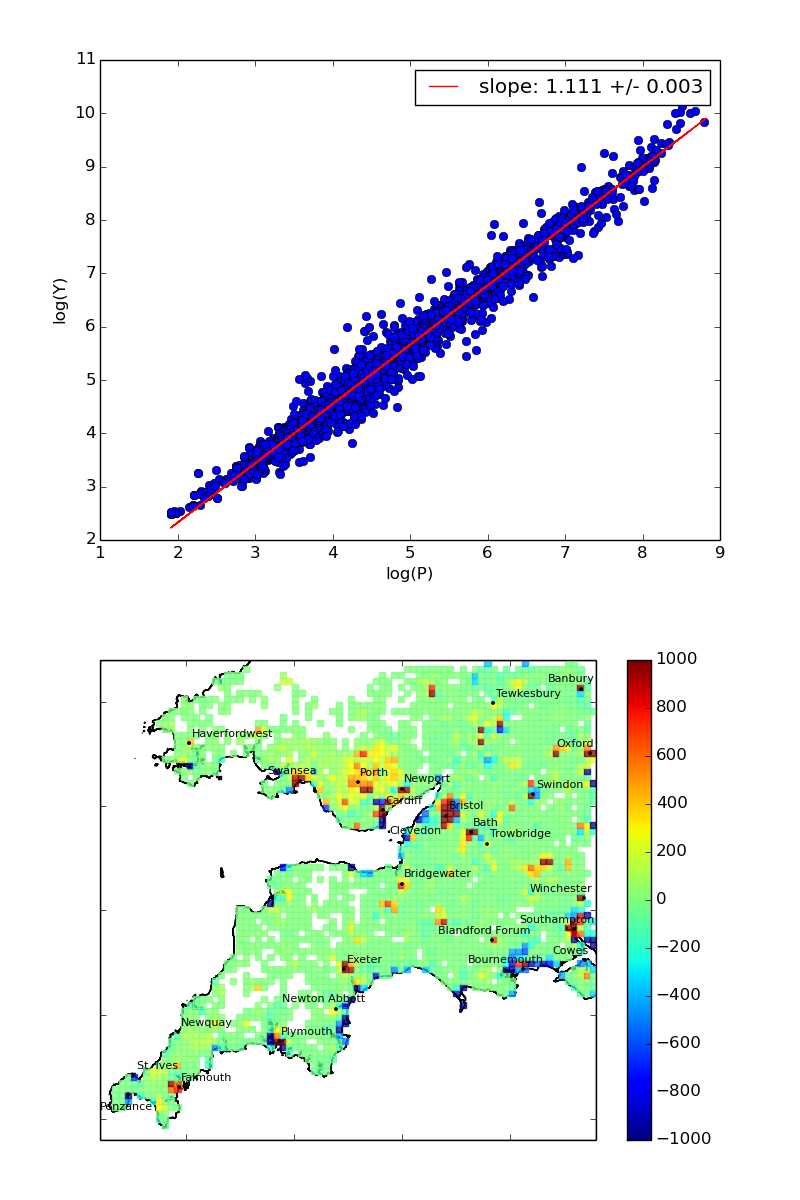}

\caption{ Top: $P$ versus $Y$ on a log-log axis. Bottom: $A_{PY}$  plotted as a heatmap over the South-West UK (with $|A_{PY}|$ capped at 1000 for better resolution). Showing the same grid boxes as Figure \ref{fig:anomaly_plots}}
\label{fig:young_plots}

\end{figure}

As previous work has shown a bias towards young people in Twitter usage, and since this information is also recorded in the census, here we attempt to relate the relative youth of a local population to deviation from the tweet density trend. Plotting the number of people aged 18 to 35 per unit area $Y$ against the population density $P$ on a log-log scale we find another power law. Figure \ref{fig:young_plots} shows the census data fit to the ansatz \begin{align}\label{eqn:yans}
Y = DP^\delta 
\end{align} This indicates that there are proportionally more young people in densely populated areas of the South-West UK. We can define the absolute and relative anomalies as before using the measured `youth' $\tilde{Y}$ and the predicted value $Y = BP^\delta$,
\begin{align}
A_{YP} &= \tilde{Y}  - DP^\delta \\ 
\hat{A}_{YP} &= \frac{  \tilde{Y}  - BP^\delta }{ \sqrt{ DP^\delta \tilde{Y}} }
\end{align}

\begin{figure}[H]
\centering 
 \includegraphics[width=\textwidth]{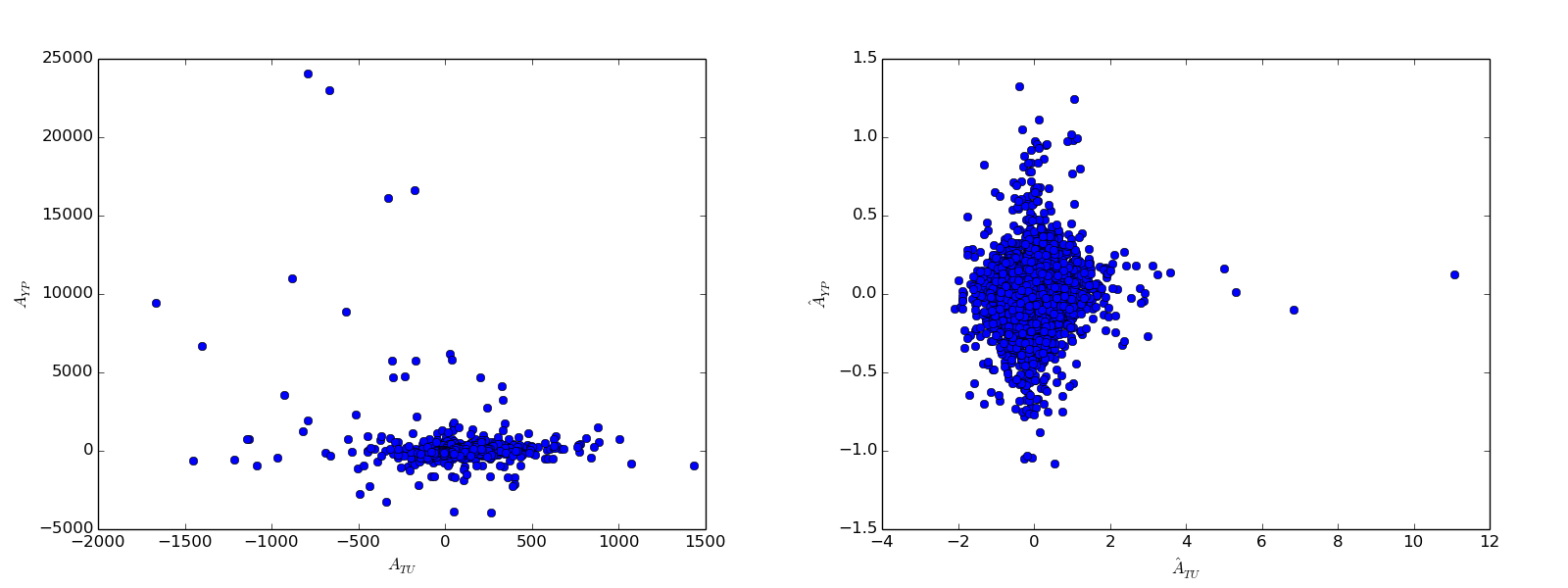}
 
\caption{ Left: Plotting absolute anomalies in young people,  $A_{YP}$,  against absolute anomaly in tweets,  $A_{TU}$. Right: Plotting relative anomalies in young people,  $\hat{A}_{YP}$,  against absolute anomaly in tweets,  $\hat{A}_{TU}$.
}
\label{fig:correlation_plots}

\end{figure}

We plot the absolute and relative anomalies for both youth and tweet density against each other in Figure \ref{fig:correlation_plots}. Neither of the observed relationships is particularly strong. The absolute anomalies have a roughly linear relationship, with many outliers; surprisingly the plot suggests a negative relationship, which would indicate that an excess of young people predicts a {\it deficit} of tweets. The relative anomalies show no clear relationship. What these plots demonstrate well is that, after accounting for the common correlate of population density, the relationship between anomalies is much weaker than the relationship between the raw quantities. We conclude that in our dataset, youth is only indicative of high Twitter activity due to the common association of both variables with population density. Demonstrating that youth (or other demographic variables, like income or ethnicity) has any additional effect beyond this is a more difficult question and we do not pursue it here.

\section{Discussion}

We have investigated relationships between population, user and tweet density.
\begin{itemize}
\item Using place-tagged tweets, we find a robust power law relationship between these quantities that holds across a range of measurement scales with the same exponent.
\item When we look at users who tweet often in our sampling area, and so probably live there, population density does a good job of predicting how many users and tweets we should see in an area.
\item Unusual areas can be identified by looking at deviations from the power law relationship and can have high or low numbers of tweets and users
\end{itemize}

Our fits of user density against population density using place-tagged tweets yield exponents, $\alpha$,$\beta$ and $\gamma$ greater than one. There is a range of distance scales over which the exponent is constant giving us some confidence that these exponents are not artefacts of the sampling procedure. Super-linear scaling laws like these are often observed for social phenomena  \cite{Tizzoni:2014} \cite{Bettencourt:2007} \cite{Schlapfer:2014}, where increasing the density of people generates synergistic effects that encourage richer and more active social, scientific and business networks. We observe the same thing here on Twitter.

We do not see the same effects when using geo-tagged tweets and have shown that geo-tagged tweets are very different in nature to place-tagged tweets. Geo-tagged tweets often originate from automated accounts or via shares from other social media platforms. Due to the large number of studies utilizing geo-tagged tweets this is an important consideration for future research.

Our study builds on previous studies which ask about sampling biases in Twitter data. Those studies consistently find a bias towards urban areas, i.e. areas with large population density. We have demonstrated that this systematic bias can be modelled with a simple power law. The number of tweets grows super-linearly with number of users, in line with work on other scaling laws in cities. Since Twitter is such a rich and useful data source we believe this study will enable better, more representative, sampling of tweets and a better interpretation of research based on Twitter. 

We have shown that, like patents or wages, the number of tweets is a function of the social environment and the number of tweets rises with population density. This is perhaps due to the denser social networks that exist in urban areas, which may make social media more attractive. The anomaly plots, Figure \ref{fig:anomaly_plots}, suggest that the social networks relevant for Twitter are not constructed {\it de novo} on Twitter itself, but reflect pre-existing social networks based in cities, towns and neighbourhoods. We can see this for cities spanning multiple grid boxes (like Cardiff, Bristol, Bournemouth etc.) the boxes overlapping the city consistently have an excess or deficit of tweets. This implies that whole cities or unified urban areas collectively make more or less use of Twitter, after accounting for population density. If Twitter had a unique social network, fully independent of external spatially embedded social networks, we would not observe these results. This suggests that the social network on Twitter is (at least partially) a manifestation of pre-existing social networks. Our findings also indicate that there are local features of some cities that make Twitter more or less attractive than is expected from population density alone. Discovering these features will be an interesting challenge for the future.

\theendnotes

\appendix

\section*{Appendix: Cross-validation}

In order to check the robustness of our fits we performed the following validation experiments. 

To check that the arbitrary bounding box does not cause artefacts we selected randomly placed boxes covering 25\% of the area of our original bounding box and located within the original sample area. We then repeated the whole analysis using only tweets in the sub-boxes, repeating this process 1000 times to construct a resampling distribution. Grid resolution was maintained, for example, when sub-sampling with $X=80$ the sub-area was covered by a $40 \times 40$ grid.

\begin{figure}[H]
\centering 
 \includegraphics[width=\textwidth]{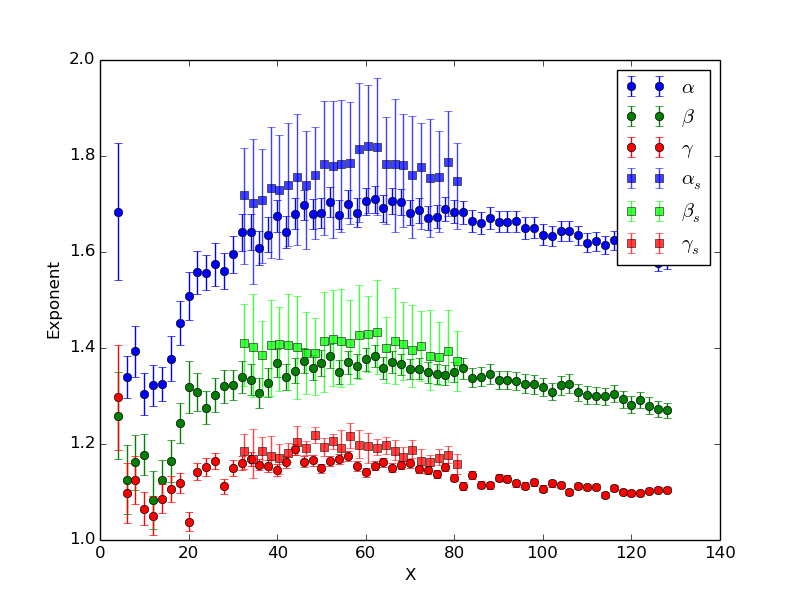}

\caption{ Fits of exponents across a range of scales. Circles show the same data as Figure \ref{fig:place_plots}, i.e. exponents fitted using all available place-tagged tweets with error bars indicating the fit uncertainty. Squares show exponents fitted from 1000 sub-samples of an area 25\% the size of the original bounding box, with error bars now showing the 68\% confidence interval of the resampling distribution. 
}
\label{fig:quadrant}

\end{figure}

We compare the sub-sampled data to the full data set in Figure \ref{fig:quadrant}. Across the scaling window, $X=32$ to $X=80$, the sub-sampled exponents are consistent with the exponents calculated from the full data set. 

\begin{figure}[H]
\centering 
\includegraphics[width=\textwidth]{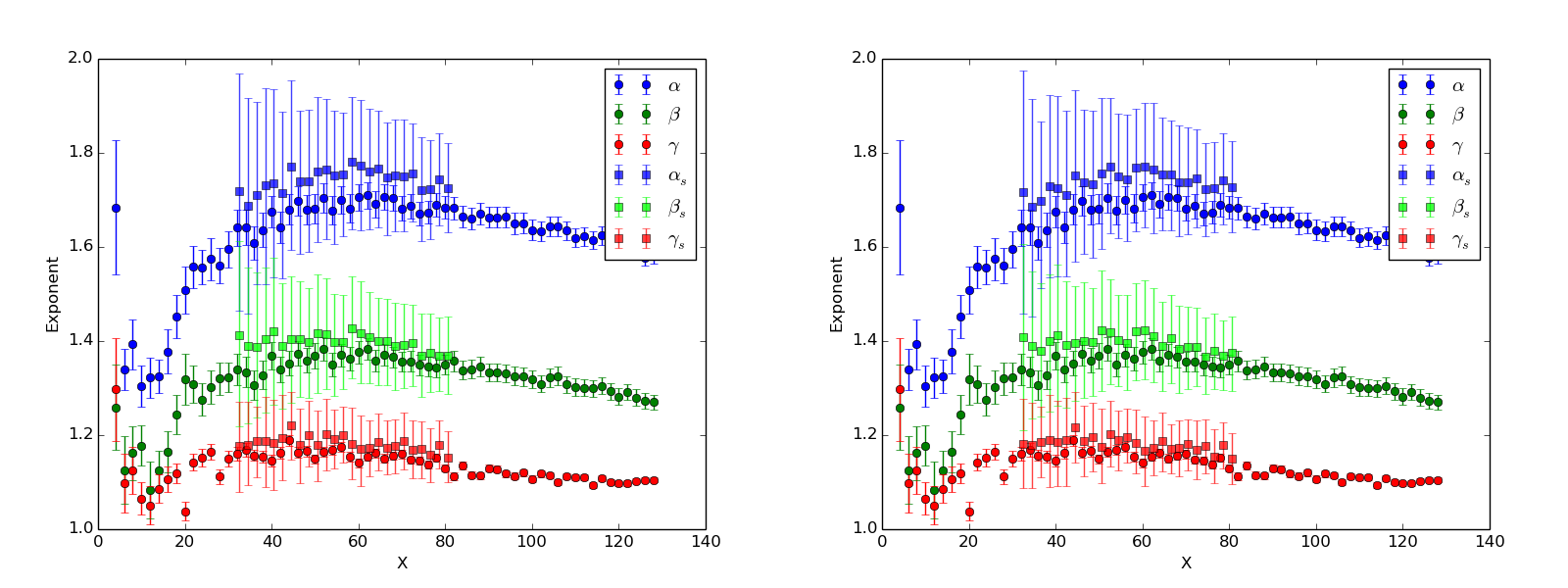}
\caption{ Fits of exponents across a range of scales. Circles show the same data as Figure \ref{fig:place_plots}, i.e. using all available tweets with error bars indicating the fit uncertainty. Squares show the same fits, using 5\% of the data, where the error bars show the 68\% confidence interval of the resampling distribution. Left: Allowing neighbouring boxes to be chosen. Right: Preventing neighbouring boxes from being chosen.
}
\label{fig:random}

\end{figure}

To check that the spatial proximity of our grid-boxes is not biasing our results, we repeated the analysis using a random subset of 5\% of the populated grid-boxes. We performed this analysis $1000$ times to create a resampling distribution for the fitted exponents. We compare the resampling distribution to the original results in Figure \ref{fig:random}. Figure \ref{fig:random} also shows the case where the random sampling is restricted to never choose two boxes sharing an edge or corner. In both cases we find the exponents are consistent with the full data set.

\end{document}